\documentclass[journal]{IEEEtran}
\usepackage{amsmath,amsfonts}
\usepackage{algorithmic}
\usepackage{algorithm}
\usepackage{array}
\usepackage[caption=false,font=normalsize,labelfont=sf,textfont=sf]{subfig}
\usepackage{textcomp}
\usepackage{stfloats}
\usepackage{url}
\usepackage{verbatim}
\usepackage{graphicx}
\usepackage{cite}
\begin{document}

\title{Using voice analysis as an early indicator of risk for depression in young adults}

\author{Klaus R. Scherer$^1$, Felix Burkhardt$^2$, Uwe D. Reichel$^2$, Florian
  Eyben$^2$, Bj\"orn W. Schuller$^{3,4,5,6,7}$\\
  $^1$Department of Psychology, University
  of Geneva, Geneva, Switzerland\\
  $^2$audEERING GmbH, Gilching, Germany\\
  $^3$Technical University of Munich, MRI, Munich, Germany\\
  $^4$University of Augsburg, Augsburg, Germany\\
  $^5$Imperial College, London, UK\\
  $^6$Munich Center for Machine Learning, Munich, Germany\\
  $^7$Munich Data Science Institute, Munich, Germany}

% The paper headers
\markboth{IEEE Transactions on Affective Computing}{}

%\IEEEpubid{0000--0000/00\$00.00~\copyright~2021 IEEE}
% Remember, if you use this you must call \IEEEpubidadjcol in the second
% column for its text to clear the IEEEpubid mark.

\maketitle

\begin{abstract}
Increasingly frequent publications in the literature report voice
quality differences between depressed patients and controls. Here, we
examine the possibility of using voice analysis as an early warning
signal for the development of emotion disturbances in young adults. As
part of a major interdisciplinary European research project in four
countries (ECoWeB), examining the effects of web-based prevention
programs to reduce the risk for depression in young adults, we
analyzed a large number of acoustic voice characteristics in vocal
reports of emotions experienced by the participants on a specific
day. We were able to identify a number of significant differences in
acoustic cues, particularly with respect to the energy distribution in
the voice spectrum, encouraging further research efforts to develop
promising non-obtrusive risk indicators in the normal speaking
voice. This is particularly important in the case of young adults who
are less likely to exhibit standard risk factors for depression such
as negative life experiences.
\end{abstract}


\section{Introduction}

Health professionals agree that the two most frequent types of mental
illness worldwide are depression and generalized anxiety disorders,
with evidence for strong co-morbidity \cite{cummings2014comorbidity,
  sartorius1996depression}. The incidence rates for these affective
disorders have risen steadily, especially during and after the COVID
pandemic.  The segment of the population most affected are adolescents
and young adults \cite{alzueta2023risk, wilson2022rising}. Given that
adolescence is a period of the life span in which rapid social,
emotional, and cognitive development and major life transitions have
to be faced, the risk for recurrent affective disorders may lead to
major impairments in interpersonal, social, educational, and
occupational functioning. Not surprisingly, health professionals give
high priority to prevention and early intervention for depression and
anxiety disorders in young people \cite{thapar2022depression}.

Consequently, a central task for research in this domain is the
identification and assessment of the major risk factors for the
development of these disorders as well as the elaboration of
appropriate programs for prevention and early intervention. The
literature shows convincing evidence for six major risk factors:
widowhood, physical abuse during childhood, obesity, having 4–5
metabolic risk factors, sexual dysfunction, and job strain (as shown
in a major umbrella review of 134 meta-analyses spanning 1283 studies;
\cite{kohler2018mapping}). In addition, stable personality traits,
such as neuroticism \cite{vittengl2017pays} and negative versus
positive trait affect \cite{dejonckheere2018bipolarity} are considered
as major risk factors for affective disorders.

As it is difficult to prevent major life events or to change major
personality dispositions and as adolescents are unlikely to have been
frequently exposed to stressful life experiences such as widowhood or
job strain, more promising targets for prevention and intervention
programs for young adults need to be identified. One potential
direction consists of identifying cognitive and emotional dysfunctions
or vulnerabilities that may affect daily life. Research in this area
has provided evidence for cognitive biases in attention, memory,
interpretation of events, associations, and ideations that can be
considered as increasing risk for emotional disturbances
\cite{mathews2005cognitive}. Specifically, it can be shown that
appraisal biases, for example, unrealistic evaluation of one's control
over events and coping with consequences of events, can create emotion
dispositions (tendencies to frequently experience anxiety or sadness)
which in turn may increase the risk for depression and anxiety
disorder \cite{mehu2015appraisal, scherer2021evidence, scherer2022appraisal}.

Cognitive biases and emotion dispositions can be identified before the
potential onset of serious and recurrent episodes of depression or
anxiety and are thus ideal candidates for early detection and
intervention in order to prevent serious clinical consequences,
especially in adolescents and young adults who may not yet have firmly
established biases or vulnerabilities. While a large number of
self-report instruments have been developed for several cognitive
biases, it is desirable to complement these with more unobtrusive
assessments of these risk factors. One particularly promising approach
is the use of voice and speech indicators for depressive states.

This approach has a long history – see \cite{scherer1987} for a review
of the early developments and classic contributions. In recent years,
research efforts have grown exponentially, partly due to the
development of sophisticated software for acoustic analysis but also
to an increased recognition of the clinical relevance of this approach
(see \cite{cummins2018speech, girard2015automated}). There are two major
review articles on the research findings on the vocal-acoustic markers
of affective disturbance, in particular depression:
\cite{cummins2015review} systematically review the source, formant,
spectral, and prosodic features that have been found to be linked to
depressive states, and, in a very recent article,
\cite{almaghrabi2023bio} review studies and highlight the effect of
depression on speech production and bio-acoustic speech
characteristics, suggesting that clinical depression diagnostics could
be augmented by machine learning based speech processing. One
potential indicator that is repeatedly reported in this literature is
that voices of individuals suffering from depression tend to show
relatively more energy in the lower frequency range of the spectrum
compared to non-depressed individuals.

However, it is notable that there is a relatively large amount of
inconsistency concerning the nature and direction of the findings as
well as on the vocal-acoustic indicators in the findings for emotional
disorders reported in this literature. One obvious reason is the
complexity of the phenomena under investigation and the number of
different factors involved. Thus, depression is a multi-faceted
clinical category with components of anhedonia, hopelessness,
helplessness, and other symptoms. And, while there is a high degree of
comorbidity between depression and generalized anxiety disorders,
\cite{cummings2014comorbidity, sartorius1996depression}, there are
some major differences with respect to the underlying symptomatology,
with respect to agitation and arousal, that are likely to have
differential effects on voice production. Another important factor
contributing to differences in findings on vocal indices of emotional
disturbances is the nature of the voice material collected (e.g.,
spontaneous vs. standard content, degree of ego-involvement, or
duration of the speech samples).

Much of this work has been done with diagnosed cases of emotion
disorders under clinical treatment in comparison to normal controls,
making it extremely difficult to obtain comparable samples and control
all pertinent variables. Furthermore, many results reported in this
literature are not directly applicable to attempts at an early
diagnosis of the risk of developing emotion disorders such as
depression and generalized anxiety, especially in younger people. The
current work explores the potential for the development of appropriate
indicators and measurement options for the development of early
detection and prevention measures in the case of risk for emotion
disorders in adolescents and younger adults. The main aim is to
identify which of a large number of standard acoustic voice variables
show a significant relationship with the central criterion for the
presence of a clinically significant risk level for depression – a
score of $\geq 10$ on the Public Health Questionnaire (PHQ-8,
\cite{kroenke2009phq}). In addition, on a more exploratory note, we
examine whether specific emotion experiences reported by the
participants for the current day are significantly related to specific
acoustic voice features and whether this relationship is mediated by
the clinical risk factor.

\section{Method}

The data reported in this article have been collected as part of a
major European Horizon 2020 project, Emotional Competence for
Well-Being in Young Adults (ECoWeB; see Acknowledgement), consisting
of web administration of appropriate diagnostic instruments and
different types of prevention packages.

\subsection{Diagnostic instruments}

To measure the risk for depression, the Public Health Questionnaire
PHQ-8 \cite{kroenke2009phq} was used. Tendencies toward Generalized
Anxiety Disorders were assessed with the GAD-7 \cite{spitzer2006brief}. In
addition, the Warwick-Edinborough Well Being scale WEMWBS
\cite{stewart2011warwick} was administered.

Emotion dispositions were assessed by a specially developed Monitor
instrument.  Users were asked to indicate which of 12 emotions they
had experienced on the current day and with what intensity. They also
had a choice of providing a detailed account of the most important
experience on that day, either by a written text or a vocal recording
of the specific situation experience. In addition, they were asked to
indicate the specific emotions they had felt in this particular
situation and rate the respective intensities.

\subsection{Voice recording}

Data collection and intervention programs in ECoWeB were administered
via phone app. In consequence, recording specifications (prompts,
microphone, conditions, sampling rate) depended on the equipment used
by the individual participant.  Due to requirements formulated by the
ethics committee of the project, the data needed to be anonymized
before being sent from the user’s device to the server for central
storage. We used random splicing (RS) \cite{scherer1971randomized,
  burkhardt2023masking}, an algorithm that cuts the sample into
segments and then re-assembles the segments in a randomized
order. This method obfuscates especially the textual content of the
samples and is well established to preserve acoustic features that
correlate with emotional expression. The software has been made open
source as part of the Nkululeko framework
\cite{burkhardt2022nkululeko}.The RS method destroys dynamic acoustic
aspects of speech. Nevertheless, we found that feature-based emotion
detection is robust against RS (see details in the Supplemental Online
Material).

\subsection{Acoustic analyses}

We used two kinds of acoustic feature sets on the data: the Nkululeko
software \cite{burkhardt2022nkululeko} and openSMILE eGeMAPS and
features extracted by the Praat software. The openSMILE framework
\cite{eyben2010opensmile} extracts acoustic feature sets that are
based on frame-based low level descriptors, such as F0 and combined by
statistical functionals. We used the eGeMAPS set
\cite{eyben2015geneva}, an expert set of 88 acoustic features. These
features work well with many speech classification tasks and have been
used as a kind of standard acoustic features in the scientific
community (about 1500 citations based on Google scholar). In addition,
we extracted typical features with the Praat software
\cite{boersma2001praat} using the scripts by David Feinberg
\cite{feinberg2019parselmouth}. Both feature sets cover various voice
quality characteristics (e.g., jitter, shimmer, harmonics to noise
ratio, spectral slope), articulatory characteristics (e.g., formant
frequencies), as well as prosodic characteristics (pitch, energy,
speech rate).

\subsection{Participants}

363 participants from four European countries (Great Britain, Belgium,
Germany, and Spain; 14.1\% male, 84.0\% female, 1.9\% other; between
16-21 years of age) recorded a vocal description of a situation that
had elicited an emotion.

165 participants recorded only one emotion experience, 198
participants recorded several situations at different days, yielding a
total of 1102 voice records. In this case, for the statistical
analyses, rated emotion intensities and acoustic parameters were
averaged over the different situations reported by single
participants.

\section{Results}

In order to determine which acoustic voice parameters show significant
differences between participants with low and high risk categorization
on the basis of their PHQ scores (cut-off level: $\geq 10$), we
computed a multivariate Analysis of CoVariance (ANCOVA), including all
acoustic parameters measured, controlling for the scores on the
Generalized Anxiety questionnaire (GAD) and the Well Being Scale
(WEMWBS) as well as for gender and country of origin. We used this
covariance model to ensure that high and low risk for Depression
scores were not affected by differential values for GAD and WEMWBS,
given that the GAD is positively and the WEMWBS negatively correlated
with PHQ, nor by gender or country differences. The results showing
the acoustic parameters with significant differences for low vs. high
risk are shown in Table \ref{tab:sigfeat}.  We ensured that these
parameters were not affected by random splicing (see details in the
Supplemental Online Material).

\begin{table*}[!t]
\caption{Acoustic parameters with significant differences ($p \leq
  .05$) between low and high risk for depression groups (PHQ $\geq
  10$) in a multivariate Analysis of CoVariance (ANCOVA, controlling
  for GAD, WEMWBS, gender, and country of origin). Note: UV =
  unvoiced, V = voiced.}
\label{tab:sigfeat}
\centering
\begin{tabular}{|c|c|c|c|c|c|}
\hline
 & Low\_Risk mean & High\_Risk mean & $p$ & Partial $\eta^2$ & Interpretation: high risk participants have: \\
\hline
F0 semitone mean & 33.77 & 32.26 & 0.005 & 0.022 & lower F0 \\
\hline
mfcc2 mean & 8.19 & 10.44 & 0.001 & 0.030 & steeper falling slope, less vocal energy \\
\hline
mfcc4 mean & -2.85 & -0.6 & 0.019 & 0.015 & hard to interpret phonetically; possibly differences \\
& & & & & in lexical content \\
\hline
logRel F0-H1-H2 mean & 7.38 & 6.16 & 0.033 & 0.013 & normalized amplitude ratio of first two harmonics; \\
& & & & & more glottal constriction (creaky voice) \\
\hline
slope V0-500 mean & 0.025 & 0.020 & 0.036 & 0.012 & less rising slope up to F1, might indicate an \\
& & & & & overall steeper falling spectral slope \\
\hline
alphaRatio UV mean & -11.51 & -12.00 & 0.021 & 0.015 & less high frequency energy in fricatives \\
\hline
F0 Hz mean & 195.76 & 183.91 & 0.004 & 0.023 & lower F0 \\
\hline
\end{tabular}
\end{table*}

These results shown in Table \ref{tab:sigfeat} suggest that the issue
of spectral balance, the relationship between the vocal energy in the
lower and upper frequency ranges, may provide information on the
likelihood of the risk for emotional disturbance. The indicators
listed in Table 1 can be subdivided into articulatory, voice quality,
and prosodic markers for high depression risk.

\paragraph{Articulation}
We find differences in alphaRatioUV mean, which in the eGeMAPS feature
set is defined as the energy of high (1-5 kHz) to low (50-1000 Hz)
energy in the spectrum in unvoiced segments. Lower values for
high-risk participants can indicate a reduced energy of high frequency
fricative noise, which can be due to a lesser degree of fricative
constriction \cite{shadle1996quantifying, barnes2020prosodic}. Along
these lines, high-risk patients overall might show an articulatory
undershoot \cite{lindblom1990explaining} and overall less energetic
speech expressed in a lower degree of articulatory constrictions.

\paragraph{Voice quality}
We find differences in logRelF0-H1-H2 as well as in the spectral slope
features mfcc2\_mean and slopeV0-500. LogRelF0-H1-H2 is defined as the
amplitude ratio of the 1st and 2nd harmonic normalized by the
amplitude of the 0th harmonic (i.e., the local spectral peak at the f0
frequency). SlopeV0-500 is the spectral slope in voiced segments
between 0 and 500 Hz which is in the first formant's frequency
region. Mfcc2\_mean denotes the arithmetic mean value of the 2nd mel
frequency cepstral coefficient.

LogRelF0-H1-H2 has been shown to be a robust correlate for glottal
constriction as given for example in creaky voice (see
\cite{keating2015acoustic} for a review). Results in Table 1 show
lower logRelF0-H1-H2 values indicating a stronger glottal constriction
with potentially more creaky voice quality for the high depression
risk participants. One might speculate that the increase in glottal
constriction, which serves to protect the trachea, serves as a defense
mechanism.

The spectral slope features mfcc2 and slopeV0-500 can be related to
vocal effort differences. Generally, increased vocal effort is
expressed in more abrupt glottal closures \cite{dromey1992glottal},
which show acoustically in an energy boost towards higher frequency
regions \cite{glave1975effort, gauffin1989spectral,
  sluijter1996spectral}, see \cite{mooshammer2010acoustic} for an
overview. This higher frequency energy boost causes a flatter spectral
slope. In line with \cite{taguchi2018major} we observe higher mfcc2
values for the high depression risk group. \cite{taguchi2018major}
show that this difference reflects reduced energy in the frequency
band from 2000 to 3000 Hz. They attribute their findings to vocal
tract changes that are not further specified. Additionally to such
potential articulatory changes, the reduced energy in this frequency
band can be interpreted to express a steeper spectral downward slope
and thus to indicate reduced vocal effort for the high depression risk
group. The second slope feature SlopeV0-500\_mean shows a less rising
slope towards the first formant’s energy peak for high depression risk
participants. This again can be attributed to an overall steeper
falling spectral slope. Thus, taken together our results indicate that
the high-risk participants can be characterized by overall less vocal
effort and a stronger glottal constriction.

\paragraph{Prosody}
High-risk participants speak with a lower pitch level
(F0semitone\_mean, meanF0Hz). This observation confirms the findings
of the majority of studies reviewed in \cite{cummins2015review}. The
lower pitch level might amongst others be attributed to overall lower
arousal (see e.g. \cite{bachorowski2008vocal}) of high risk
participants.

Finally, mfcc4\_mean, the mean value of the fourth Mel frequency
coefficient differs between the high and low risk groups which
confirms the findings of \cite{zhao2022vocal}, who observed a
significant relation between this coefficient and the PHQ-9
score. This feature is harder to interpret in phonetic terms as it
might indicate differences in the lexical content in the emotion
reports of low and high risk participants that can be observed
regardless of the spoken language.

In a second step, we ran a Discriminant analysis with the binary scale
PHQ risk (high -- low) as the dependent variable and the set of
acoustic parameters yielding significant differences in the ANCOVA
analysis described above, as predictors.  Table \ref{tab:discranalysis} shows the structure
matrix of the analysis.  66.4\% (cross-validated 63.9\%) of original
grouped cases were correctly classified (Wilks lambda 0.952, p <
0.015). This analysis confirms the role of spectral balance as a
potential indicator of risk for depression, suggesting that a
classification rate that is somewhat better than pure chance can be
achieved.

\begin{table}[!t]
\caption{Discriminant analysis of high vs. low risk for depression
  based on the variable set reaching significance in the ANCOVA shown
  in Table \ref{tab:sigfeat}. Note: Pooled within-groups correlations
  between discriminating variables and the standardized canonical
  discriminant function. Variables ordered by absolute size of
  correlation within function Classification 66.4\% (cross-validated
  63.9\%) of original grouped cases correctly classified (Wilks lambda
  $0.952$, $p < 0.015$).}
\label{tab:discranalysis}
\centering
\begin{tabular}{|c|c|}
\hline
Structure Matrix  & Function 1 \\
\hline
mfcc4 mean & $.576$ \\
\hline
mfcc2 mean & $.558$ \\
\hline
logRelF0-H1-H2 mean & $-.548$ \\
\hline
F0semitoneFrom mean & $-.533$ \\
\hline
meanF0Hz & $-.529$ \\
\hline
slopeV500-1500 mean & $-.411$ \\
\hline
alphaRatioUV mean & $-.229$ \\
\hline
\end{tabular}
\end{table}

As mentioned on the outset, we also wanted to explore the possibility
of using vocal cues to identify emotion dispositions, which have been
shown to play a role in the development of risk for depression. A
major theoretical prediction on the origin of risk for depression is
the assumption that an appraisal bias of low control/coping ability
tends to produce an emotion disposition for experiencing frequently
anxiety and sadness \cite{scherer2021evidence,
  scherer2022appraisal}. Using an Emotion Monitor instrument, the
participants in the current study reported on each assessment day with
which intensity (from 0 to 7) they had experienced each 12 emotions,
seven of which are considered as particularly relevant for depression
risk: high anxiety, sadness, shame, and low amusement, joy,
pleasure. A discriminant analysis on high/low risk with stepwise entry
showed that high intensity of anxiety was the best predictor for risk
(sig. of F to remove .044). To a lesser extent low intensity of joy
also contributed (sig. of F to remove .063).

\section{Discussion}

As a limitation there are unequal numbers of participants in the high
and low risk group, which affects significance testing, but this is
unavoidable as persons at risk will always constitute a smaller
portion of an unselected sample. There is a biased gender distribution
(84\% female to 14\% male), reflecting the overall bias in the ECoWeB
study as a whole. Furthermore, participants in the study could choose
whether to use the option to vocally report an emotion experience on
each one of the several survey dates, but only a small percentage of
the full sample did so.

\section{Conclusion}

As outlined in the Results section we found articulatory, prosodic, as
well as voice quality markers for high depression risk: Articulatory
features indicate an articulatory undershoot and overall less
energetic speech expressed in a lower degree of fricative
constrictions. Prosodically, we mainly observe a lower pitch register
related to lower arousal showed as an indicator for high depression
risk. Voice quality markers suggest an overall reduction in vocal
effort as well as a tighter glottal constriction. These findings
confirm and elaborate earlier results on the important role of
acoustic indicators of depressive disorder, particularly with respect
to the importance of the mel frequency cepstral coefficient (mfcc)
parameters.

Given that we found these significant effects in a study of potential
risk factors with young adolescent participants not currently
diagnosed with clinical depression but passing the established
threshold for corresponding risk on an established diagnostic
instrument, indicates the utility of investing in further research
using voice analysis to determine risk factors for emotional disorders
in a timely fashion. While it is unlikely that acoustic parameters
alone can guarantee a valid diagnosis of risk for depression, they can
be most valuable in addition to other behavioral indicators (such as
facial expression, see \cite{girard2015automated} and self report
indices such as appraisal bias and negative emotion dispositions
\cite{scherer2022appraisal}.

\section{Acknowledgments}

This work was conducted as part of the Emotional Competence for
Well-Being in the Young (ECoWeB) Consortium. ECoWeB is a team of
researchers across European higher education institutions
(Universities of Exeter, Oxford, Ghent, LMU Munich, Jaume I, Glasgow,
Essex, Geneva, Bern; Panteion University of Social and Political
Sciences) and companies (audEERING; Monsenso) funded by the European
Union’s Horizon 2020 research and innovation funding scheme
(SC1-PM-07-2017; grant no. 754657). This consortium designed the
ECoWeB cohort and trials, prepared the digital interventions and
measures, recruited and followed up eligible young people in the UK,
Germany, Spain and Belgium, and compiled and analysed the collected
data. The members of the consortium (in alphabetical order) are:

\begin{itemize}
\item Benjamin Aas (Department of Psychology, LMU Munich, Germany)
\item Holly Bear (Department of Psychiatry, University of Oxford, UK)
\item Cristina Botella (Universitat Jaume I, Spain; CIBER
  Fisiopatología Obesidad y Nutrición (CIBERObn), Instituto Salud
  Carlos III, Spain)
\item Felix Burkhardt (audEERING, Germany)
\item Timothy Cranston (Faculty of Health \& Life Sciences, University
  of Exeter, UK)
\item Thomas Ehring (Department of Psychology, LMU Munich, Germany)
\item Mina Fazel (Department of Psychiatry, University of Oxford, UK)
\item Johnny R.J. Fontaine (Department of Work, Organization and
  Society, Ghent University, Belgium)
\item Mads Frost (Monsenso A/S, Copenhagen, Denmark)
\item Azucena Garcia-Palacios (Universitat Jaume I, Spain; CIBER
  Fisiopatolog\'{i}a Obesidad y Nutrici\'{o}n (CIBERObn), Instituto Salud
  Carlos III, Spain)
\item Ellen Greimel (Department of Child and Adolescent Psychiatry,
  Psychosomatics and Psychotherapy, University Hospital, LMU Munich,
  Germany)
\item Christiane H\"o\"sle (Department of Psychology, LMU Munich, Germany)
\item Arpi Hovasapian(Department of Work, Organization and Society,
  Ghent University, Belgium)
\item Claire Hulme (Faculty of Health \& Life Sciences, University of
  Exeter, UK)
\item Veerle E.I. Huyghe (Department of Work, Organization and
  Society, Ghent University, Belgium)
\item Nanna Iversen (Monsenso A/S, Copenhagen, Denmark)
\item Kostas Karpouzis (Panteion University of Social and Political
  Sciences, Greece)
\item Johanna L\"ochner (Department of Psychology, LMU Munich, Germany;
  Department of Child and Adolescent Psychiatry, Psychosomatics and
  Psychotherapy, University of T\"ubingen, Germany)
\item Guadalupe Molinari (Universitat Jaume I, Spain; CIBER
  Fisiopatología Obesidad y Nutrición (CIBERObn), Instituto Salud
  Carlos III, Spain)
\item Alexandra Newbold (Mood Disorders Centre, School of Psychology,
  University of Exeter, United Kingdom)
\item Reinhard Pekrun (Department of Psychology, University of Essex,
  UK, and Institute for Positive Psychology and Education, Australian
  Catholic University, Sydney, Australia)
\item Belinda Platt (Department of Child and Adolescent Psychiatry,
  Psychosomatics and Psychotherapy, University Hospital, LMU Munich,
  Germany)
\item Tabea Rosenkranz (Department of Psychology, LMU Munich, Germany)
\item Klaus R. Scherer (University of Geneva, Switzerland)
\item Katja Schlegel (University of Bern, Switzerland) 
\item Bj\"orn W. Schuller (audEERING, Germany; Embedded Intelligence
  for Health Care and well-being, University of Augsburg; Group on
  Language, Audio \& Music, Imperial College London)
\item Gerd Schulte-Korne (Department of Child and Adolescent
  Psychiatry, Psychosomatics and Psychotherapy, University Hospital,
  LMU Munich, Germany)
\item Carlos Suso-Ribera (Universitat Jaume I, Spain; CIBER
  Fisiopatología Obesidad y Nutrici\'{o}n (CIBERObn), Instituto Salud
  Carlos III, Spain)
\item Rod S. Taylor (MRC/CSO Social and Public Health Sciences Unit \&
  Robertson Centre for Biostatistics, School of Health and Well Being,
  University of Glasgow, UK)
\item Varinka Voigt (Department of Psychology, LMU Munich, Germany)
\item Maria Voss (Department of Psychology, LMU Munich, Germany)
\item Fiona C. Warren (Faculty of Health \& Life Sciences, University
  of Exeter, UK)
\item Edward R. Watkins (Mood Disorders Centre, School of Psychology,
  University of Exeter, United Kingdom; Faculty of Health \& Life
  Sciences, University of Exeter, UK)
\end{itemize}

\bibliographystyle{IEEEtran}
\bibliography{references}

\newpage
\section{Biography Section}

\begin{IEEEbiography}[{\includegraphics[width=1in,height=1.25in,clip,keepaspectratio]{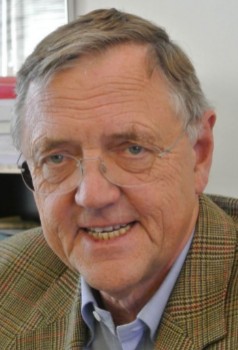}}]{Klaus R. Scherer}
Klaus Scherer (Ph.D. Harvard University) has held professorships at the University of Pennsylvania and the Universities of Kiel, Giessen, and Geneva. He is currently an emeritus professor at the University of Geneva and an honorary professor at the University of Munich. His extensive work on different aspects of emotion, in particular vocal and facial expression and emotion induction by music, has been widely published in international peer-reviewed journals. Klaus Scherer is a fellow of several international scientific societies and a member of several learned academies. He founded and directed the Swiss Center for Affective Sciences, held an Advanced Grant of the European Research Council and has been awarded honorary doctorates by the universities of Bologna, Bonn, and Fribourg. 
\end{IEEEbiography}
  
\begin{IEEEbiography}[{\includegraphics[width=1in,height=1.25in,clip,keepaspectratio]{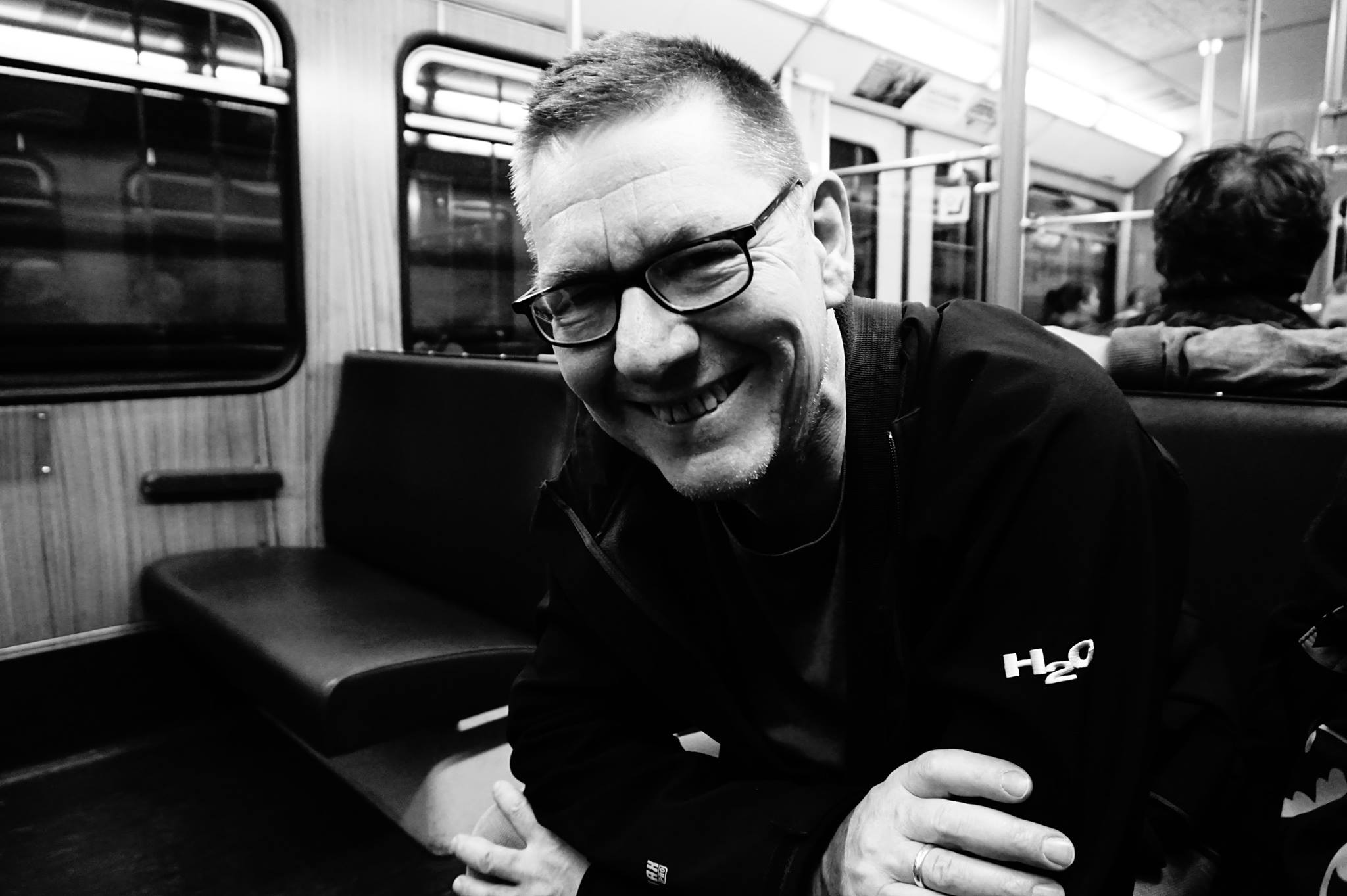}}]{Felix Burkhardt}
 Dr. Felix Burkhardt does teaching, consulting, research and development with respect to speech based emotional human-machine interfaces.
Originally an expert of Speech Synthesis at the Technical University of Berlin, he wrote his ph.d. thesis on the simulation of emotional speech by machines, recorded the Berlin acted emotions database, "EmoDB", and maintains several open source projects, including the emotional speech synthesizer Emofilt, the speech labeling and annotation tool Speechalyzer and the machine learning platform Nkululeko. Since 2018 he is the research director at audEERING after having worked for the Deutsche Telekom AG for 18 years. From 2019 to 2022 also was a full professor at the institute of communication science of the Technical University of Berlin.. 
\end{IEEEbiography}

\begin{IEEEbiography}[{\includegraphics[width=1in,height=1.25in,clip,keepaspectratio]{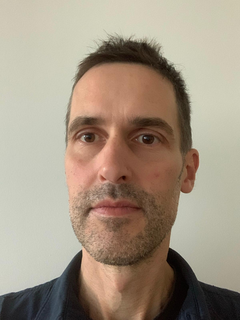}}]{Uwe D. Reichel}
Uwe D. Reichel is working as a senior researcher at audEERING GmbH and
is also affiliated at the Hungarian Academy of Sciences.  He received
his master of science and his doctoral degree in Speech Science at the
University of Munich in 2002 and 2010. The topic of his doctoral
thesis was automatized intonation modelling and its linguistic
interpretation. He worked as a research fellow at the Institute of
Phonetics and Speech Processing, Munich, and as a Humboldt research
fellow at the Institute for Linguistics at the Hungarian Academy of
Sciences in Budapest, the latter financed by a 2 years Feodor Lynen
research grant by the Alexander von Humboldt foundation from 2015 till
2017. His work covers speech- and text-based modeling and feature
engineering for various paralinguistic and digital health topics.
\end{IEEEbiography}

\begin{IEEEbiography}[{\includegraphics[width=1in,height=1.25in,clip,keepaspectratio]{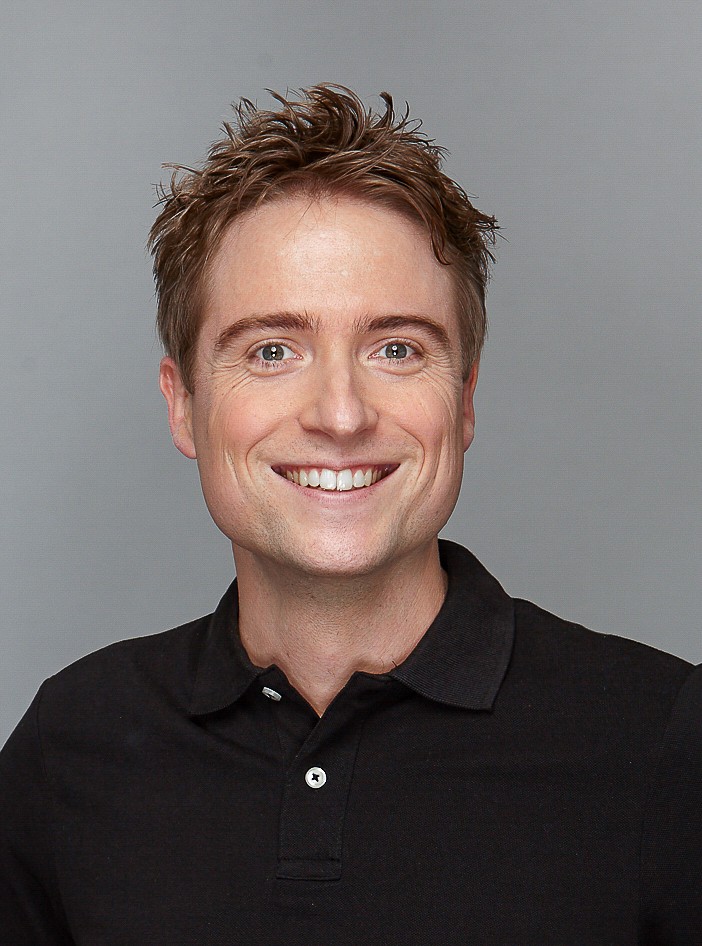}}]{Florian Eyben}
Leading tech and innovation at audEERING to deliver world leading products for speech emotion recognition and Deep Learning based audio analysis. Did a PhD at TUM, Munich, Germany on Computational Paralinguistics; expert in deep learning, audio feature extraction, signal processing, project management and tech innovation; lead author of the openSMILE toolkit, co-author of the GPU accelerated LSTM-RNN training toolkit CuRRENNT.
\end{IEEEbiography}
  
\begin{IEEEbiography}[{\includegraphics[width=1in,height=1.25in,clip,keepaspectratio]{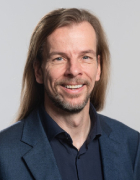}}]{Bj\"orn W. Schuller}
Björn W. Schuller received his diploma, doctoral degree, habilitation, and Adjunct Teaching Professor in Machine Intelligence and Signal Processing all in EE/IT from TUM in Munich/Germany where he is Full Professor and Chair of Health Informatics. He is also Full Professor of Artificial Intelligence and the Head of GLAM at Imperial College London/UK, co-founding CEO and current CSO of audEERING amongst other Professorships and Affiliations. Previous stays include Full Professor at the University of Augsburg/Germany and University of Passau/Germany, Key Researcher at Joanneum Research in Graz/Austria, and the CNRS-LIMSI in Orsay/France. He is a Fellow of the ACM, Fellow of the IEEE and Golden Core Awardee of the IEEE Computer Society, Fellow of the BCS, Fellow of the ELLIS, Fellow of the ISCA, Fellow and President-Emeritus of the AAAC, and Elected Full Member Sigma Xi. He (co-)authored 1,500+ publications (60,000+ citations, h-index 115), is Field Chief Editor of Frontiers in Digital Health, Editor in Chief of AI Open and was Editor in Chief of the IEEE Transactions on Affective Computing amongst manifold further commitments and service to the community. His 50+ awards include having been honoured as one of 40 extraordinary scientists under the age of 40 by the WEF in 2015. Currently, he was awarded ACM Distinguished Speaker for the term 2024-2027 and IEEE Signal Processing Society Distinguished Lecturer 2024.
\end{IEEEbiography}

\newpage
\appendix 
\section{Validation of Random Splicing}

\begin{figure}[!t]
\centering
\includegraphics[width=3.5in]{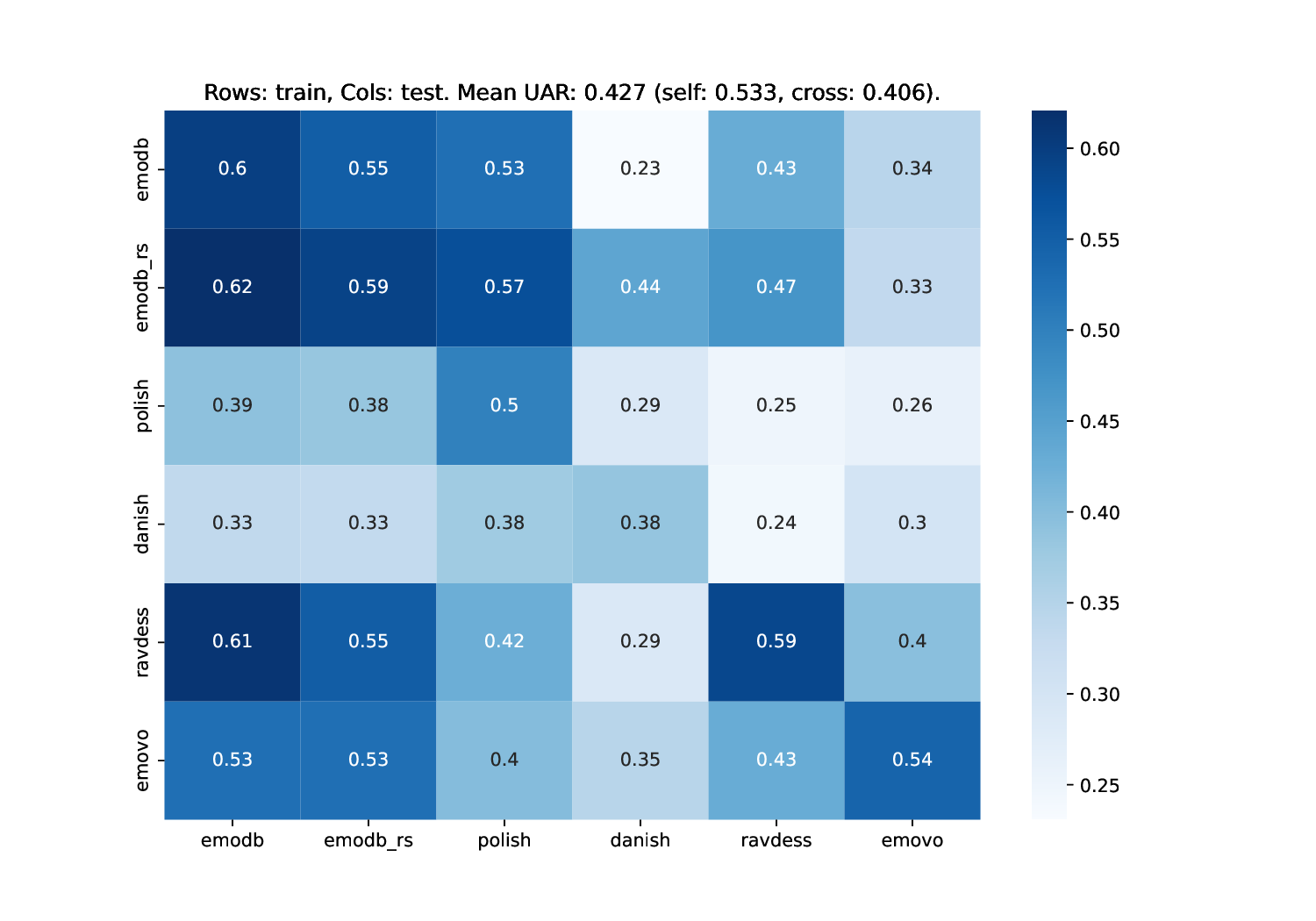}
\caption{Cross dataset emotion prediction performances to assess the
  impact of random splicing of the the Emo-DB training set (emodb\_rs) on model
  performance.}
\label{fig:randomsplicing}
\end{figure}

Because the Random Splicing (RS) method destroys the dynamic contours of the speech
samples, we need to check whether central acoustic parameters that
correlate with the prediction of emotional speech are disrupted. To
test this, we performed RS on a whole database, namely the Berlin
Emo-DB \cite{burkhardt2005database}. This database contains speech
sampled from 10 German actors portraying 6 emotion categories plus a
neutral version. We then compared this version with the original
database as well as four other international databases, namely a
Polish database \cite{powroznik2017}, the US-American Ravdess database
\cite{livingstone2018}, the Italian Emovo database
\cite{costantini2014emovo} and the Danish emotion database
\cite{engberg1997design}. All of these include samples with ground-truth
labels for the emotional categories neutral, happy, angry and sad. We
performed 36 ($6 \cdot 6$) machine learning experiments with these,
using each database once, in a cross database setting, as a test
split, once as a training split, and once doing the mono-database
experiment (using the own train and test splits). As acoustic features
we used the same that are the basis of this paper's analysis: The
eGeMAPS set \cite{eyben2015geneva}, an expert set of 88 acoustic
features and the Praat software \cite{boersma2001praat} features,
using the scripts by David Feinberg
\cite{feinberg2019parselmouth}. Both feature sets simply got
concatenated. As a classifier we used the Support Vector Machine (SVM)
algorithm implemented by the sklearn package with a misclassification
cost of $.1$. We did no optimization, as we were interested only in
the comparison between experiments. The results are shown in Figure
\ref{fig:randomsplicing}. It shows that the RS version of Berlin Emo-DB
performs as good (and in parts even better) as the original one.

\begin{figure}[!t]
\centering
\includegraphics[width=3in]{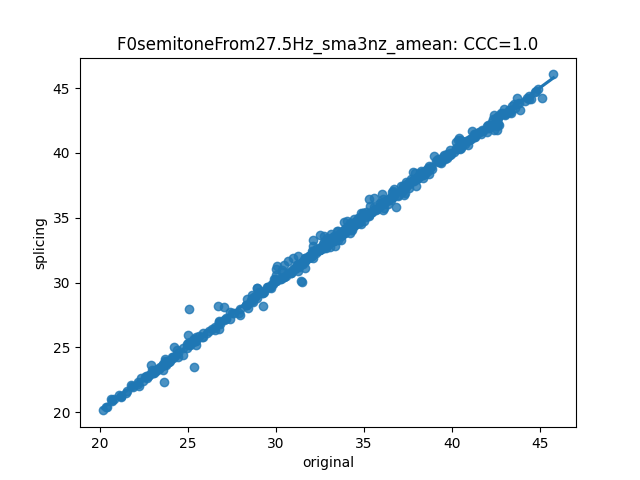}
\caption{Correlation of mean f0 values calculated on original vs
  randomly spliced audio files. The robustness of this feature against
  random splicing is indicated by a high correlation along the
  diagonal.}
\label{fig:robustfeat}
\end{figure}

Furthermore, we ensured, that the acoustic parameters with significant
differences for low vs. high depression risk shown in Table
\ref{tab:sigfeat} were not affected by random splicing, the following
way: we extracted the parameters on the Emo-DB dataset
\cite{burkhardt2005database} on the original as well as on the randomly
spliced signals and measured the concordance correlation coefficient
(CCC) between the two variants. We defined feature robustness in terms
of a CCC higher than .95. All significant acoustic parameters reached
a CCC of .99 or higher indicating that they are not or only negligibly
affected by random splicing. Figure \ref{fig:robustfeat} shows an
example correlation plot for the robust f0 mean feature.

\end{document}